\documentclass[prb,aps,floats,amssymb,showkeys,showpacs,
superscriptaddress]{revtex4}
\usepackage{graphicx}
\usepackage{graphics}
\usepackage{amsmath}
\usepackage{epsfig,bm}
\usepackage{rotating}
\usepackage{float} 
\begin{document}

\title{Positivity of the virial coefficients in lattice dimer models
and upper bounds on the number of matchings on graphs}

\author{P. Butera}
\email{paolo.butera@mib.infn.it}
\affiliation
{Dipartimento di Fisica Universita' di Milano-Bicocca\\
and\\
Istituto Nazionale di Fisica Nucleare \\
Sezione di Milano-Bicocca\\
 3 Piazza della Scienza, 20126 Milano, Italy}

\author{P. Federbush}
\email{pfed@umich.edu}
\affiliation
{Department of Mathematics\\
University of Michigan \\
Ann Arbor, MI 48109-1043, USA\\}

\author{M. Pernici} 
\email{mario.pernici@mi.infn.it}
\affiliation
{Istituto Nazionale di Fisica Nucleare \\
Sezione di Milano\\
 16 Via Celoria, 20133 Milano, Italy}

\date{\today}
\begin{abstract}
  Using a simple relation between the virial expansion coefficients of
  the pressure and the entropy expansion coefficients in the case of
  the monomer-dimer model on infinite regular lattices, we have shown
  that, on hypercubic lattices of any dimension, the virial
  coefficients are positive through the 20th order.  We have observed
  that all virial coefficients so far known for this system are
  positive also on infinite regular lattices with
  different
  structure. We are thus led to conjecture that the virial expansion
  coefficients $m_k $ are always positive.

  These considerations can be extended to the study of related bounds
  on  finite graphs generalizing the infinite regular lattices,
  namely the finite grids and the regular biconnected graphs.  The
  validity of the bounds $\Delta^k {\rm \ln}(i! N(i)) \le 0$ for $k \ge
  2$, where $N(i)$ is the number of configurations of $i$ dimers on
  the graph and $\Delta$ is the forward difference operator, is shown
  to correspond to the positivity of the virial coefficients.

  Our tests on many finite lattice graphs indicate that on large
  lattices these bounds are satisfied,
  giving  support to the  
  conjecture on the positivity of the virial coefficients.
 Moreover, in an exhaustive survey of some classes of regular
  biconnected graphs with a not too large number $v$ of vertices, we
  observe only few violations of these bounds.
  We conjecture that the frequency of the violations vanishes as $v \to
  \infty$.

  Using an inequality by Heilmann and Lieb, we find rigorous upper
  bounds on $N(i)$ valid for arbitrary graphs and for regular graphs.
  The similarity between this inequality and the one conjectured above
  suggests that one study the stricter inequality $m_k \ge \frac{1}{2k}$ 
  for the virial coefficients, which is valid for all the
  known  coefficients of the infinite regular lattice models.

\end{abstract}
\pacs{ 05.50.+q, 64.60.De, 75.10.Hk, 64.70.F-, 64.10.+h}
\keywords{Dimer problem, graph entropy, upper bounds on matchings in
regular graphs } \maketitle

\section{Introduction}
The virial series-expansion coefficients of the pressure have been computed
through relatively high orders for the monomer-dimer (MD) models on
various infinite regular lattices.  Presently, they are
tabulated[\onlinecite{kenzie}] through order $19$ for the tetrahedral
lattice and through order $7$ for the hexagonal
lattice[\onlinecite{Ftri}].  Moreover, they are known for the
triangular and the face-centered-cubic lattices through the orders
$14$ and $10$ respectively[\onlinecite{gaunt, kurtze}].  In the case
of the linear lattice[\onlinecite{fisher, fp}] and of the Bethe
lattice[\onlinecite{nagle, stil}], they are all known.  We have recently
computed[\onlinecite{bp1}] the expansions through the order $24$ for the
$bcc$ lattices in $d=3,4,5,6,7$, and for the (hyper)-simple-cubic
lattices, through order $24$ in the case of the square, cubic and $4d$
lattices, through the orders $22$ and $21$ in dimensions $d=5$ and $
6$ respectively, and finally[\onlinecite{bfp}] in general dimensions $d>6$
through the order $20$.

Long ago, Heilmann and Lieb[\onlinecite{HL1, HL}] have studied the MD
models also on finite graphs. They have shown that the zeroes of the
matching generating polynomial  $M(z)= \sum N(i) z^i$ of a graph lie
only on the real negative axis of the complex $z$ plane. For the MD
gas on a finite lattice, this implies the absence of any
phase-transition in the thermodynamic limit.

 In this paper, we show that the first $20$ coefficients of the virial
 expansion on hypercubic infinite regular lattices of any dimension are
 positive.  With the knowledge that all the coefficients
 computed so far for the virial expansions of the MD models
 on all infinite regular lattices are positive, we are led 
  to {\it conjecture} that {\it all} the virial coefficients  are
 positive for the infinite regular lattice models.

Using the definition of the graph dimer entropy[\onlinecite{bfppos}],
we argue that for a finite regular graph the bounds which correspond
to the positivity of the virial coefficients $m_k$ for infinite regular lattices are
\begin{equation}
\Delta^k {\rm \ln}(i! N(i)) \le 0
\label{Delta0}
\end{equation}
with $k=2,...,\nu$ and $i=0,...,\nu-k$, where $\nu$ is the matching
number of $G$, i.e. the maximum number of pairwise disjoint edges of
$G$.  

We have systematically tested the validity of Eq. (\ref{Delta0}) 
within two classes of finite graphs related to infinite regular 
lattices.  The first class consists of the graphs induced by finite 
grids, the second consists of the biconnected regular simple graphs.
The tests of Eq. (\ref{Delta0}) for the latter class have been
performed by exhaustively generating the biconnected regular graphs
having a not too large number of vertices $v$, with the aid of the
{\it Nauty} program[\onlinecite{nauty}] via the {\it Sage}
interface[\onlinecite{Sage}].

The results of our survey are summarized in a set of schematic tables
reported in the fourth Section. We have firstly restricted our
attention to the class of finite lattices, being interested in
finding sequences of finite lattices with no violations of the bounds
in Eq. (\ref{Delta0}).  This would indicate that all the virial
coefficients are positive for the corresponding infinite regular
lattices.  Even though the bounds in Eq. (\ref{Delta0}) have been derived
only for regular graphs, it is however interesting to consider them
also in the case of finite grids with open boundary conditions ($bc$),
(which clearly are not regular on the boundary).

In summary,  we found  violations only for ``narrow'' lattices.
This fact  supports the conclusion that  the bounds are not violated
 in the limit of infinite lattices.

We then turned to the more general class of the bipartite biconnected
 regular simple graphs. For the 
$3$-regular graphs, we were able to test these bounds for $v \le 30$
vertices. We have observed a few violations of the bounds for $v \ge
18$, but the frequency of these violations decreases
regularly for $v \ge 18$.

In the case of the bipartite biconnected $4$-regular graphs, we have tested
Eq. (\ref{Delta0}) for $v \le 22$. We have observed a single violation
for $v=20$. The frequency of the violations is even lower for $v=22$.

In the case of the bipartite biconnected  $r$-regular graphs with $r>4$,  
we could carry our tests only for $ v \le 20$, finding a single violation, 
 that occurs in a   $5$-regular graph.

When considering the non-bipartite biconnected $3$-regular graphs, we
observed few violations for $v \le 22$, starting from $v=12$; again
the frequency of the violations decreases regularly after the first
violation, for even values of $v \ge 12$. Notice that there are no such graphs
with $v$ odd.

In the case of the non-bipartite biconnected $4$-regular graphs, the 
frequency of the violations decreases regularly for even $v$
after the first violation and a similar trend is observed for odd $v$.
We could complete the tests only for $v \le 17$.  In this case 
 one has to consider over $80$ million graphs.

Based on the results of our survey, we are led to {\it conjecture}
that, for biconnected regular graphs, the bounds are violated with a
frequency that vanishes as $v \to \infty$.

Let us now observe that the validity of the bounds of Eq. (\ref{Delta0})
in the case $k = 2$ follows from the Heilmann-Lieb inequality Eq. (4)
in [\onlinecite{HL1}].
Using this Heilmann-Lieb inequality, we shall derive rigorous upper
bounds for $N(i)$ of general graphs. In the case of regular graphs
they are stricter than those in [\onlinecite{fried}] in a region with
small dimer density; in the case of
general graphs, we obtain an upper bound for the matching generating
polynomial which is stricter  than the one found in [\onlinecite{carroll}].

The (rigorous) Heilmann-Lieb inequality has the form
\begin{equation}
\Delta^k {\rm \ln}(i!(n-i)!N(i)) \le 0
\label{hlk}
\end{equation}
 with $n=[\frac{v}{2}]$ and $k=2$. The similarity between this
inequality and Eq. (\ref{Delta0}) led us to investigate Eq. (\ref{hlk})
also for $k \ge 2$ and to {\it conjecture} for infinite regular
lattices the stricter bound $m_k \ge \frac{1}{2k}$, which indeed is
satisfied by all known virial coefficients.  As in the previous case,
we have investigated how well the bounds Eq. (\ref{hlk}), with $k\ge
2$, are satisfied by finite grids and regular biconnected graphs.

The tests of Eq. (\ref{hlk}), with $k\ge 2$, on lattice graphs give
 more violations than for Eq. (\ref{Delta0}).

Therefore the examination of these lattices still gives some indication, though
not as clear as in the case of Eq. (\ref{Delta0}), that the bounds
Eq. (\ref{hlk}), with $k\ge 2$, are satisfied by the corresponding infinite
lattices and that the virial coefficients are positive.

For the general graphs produced with the aid of {\it Nauty}, we find similar
results as with the bounds of Eq. (\ref{Delta0}), although we observe more 
violations.

We have also shown that, for any $v$, the bounds of Eqs. (\ref{Delta0},
\ref{hlk}) are satisfied by the  approximate distribution of bipartite regular
random-graph  obtained in [\onlinecite{FKM}]. In the case of
bipartite regular biconnected graphs, we argue that a
conjecture on the entropy for these graphs, made in [\onlinecite{FKM, FKM1}],
implies the conjecture that for $v \to \infty$ the frequency of the violations 
tends to zero.

In all the tests performed on regular biconnected graphs (over more than $300$
million graphs), Eq. (\ref{hlk}) is valid for $k=3$.  We shall discuss
also the corresponding upper bounds on $N(i)$;
the examples considered indicate that
in the case of bipartite regular biconnected graphs they are less strict than
the Upper Matching conjecture (UMC) [\onlinecite{FKM}], when the latter applies.

Let us mention that while the UMC conjecture has not been yet proven,
the lower asymptotic matching conjecture stated in [\onlinecite{FKM}]
has been proven in [\onlinecite{Csi}] and [\onlinecite{GS}].

The paper is organized as follows. In the second Section, using a
simple relation between the coefficients of the virial expansion of
the pressure and the expansion coefficients of the dimer entropy, we
prove that on hypercubic lattices the virial coefficients through
order $20$ are positive for generic $d$.  In the third Section, we
derive Eq. (\ref{Delta0}),  obtain rigorous upper bounds for $N(i)$ and
discuss Eq. (\ref{hlk}).
The fourth Section summarizes the results
of the graph tests for a variety of lattices and graphs.  In Appendix
A we give the virial coefficients for $d=2,3$ through order $24$. In Appendix
B, the validity of Eq. (\ref{Delta0}), and of Eq. (\ref{hlk})
is proven for a few classes of graphs and for two average distributions.
In Appendix C a formula for the bound on $N(i)$ is deduced from 
the inequalities of Eq. (\ref{hlk}).

\section{Positivity of virial coefficients} 

The combinatorial-statistical properties of a MD system on a $r$-regular 
lattice are usually described in the grand-canonical formalism of 
statistical mechanics, in which the pressure is defined as
\begin{equation}
\lim_{v \to \infty} \frac{1}{v} {\rm \ln} \Xi_v(z)  =
 P(z) =\sum_i b_iz^i. 
\label{pr}
\end{equation}
 Here $\Xi_v(z)$ is the grand-canonical partition function for a
 $v$-site lattice and $z$ is the activity.  The dimer
 density is then
\begin{equation}
\rho(z) = z \frac{dP} {dz} = \sum_{i=1}^{\infty}i b_i z^i
\label{den}
\end{equation}

Setting $p = 2\rho$, and solving Eq. (\ref{den}) for $z = z(p)$,
we can express the pressure in terms of $p$
\begin{equation}
P(p) = p/2 + \sum_{k=2}^{\infty} m_k p^k
\label{vir}
\end{equation}
This is the virial expansion.
The entropy is defined by
\begin{equation}
\lambda(p)=-\rho(z) {\rm \ln} z + P(z)
\label{entr}
\end{equation}
from Eqs.(\ref{den}) and (\ref{entr}) one gets[\onlinecite{fp2, bfp}]
\begin{equation}
\frac{d \lambda}{d p} = -\frac{\ln z}{2}
\label{dla}
\end{equation}
Using the expansion[\onlinecite{FF, bfp}]
\begin{equation}
\lambda(p)= R(p) + \sum_{k=2}^{\infty}a_k p^k
\label{entr2}
\end{equation}
where
\begin{equation}
R(p) = \frac{1}{2}(p{\rm \ln}(r)-p{\rm \ln}p -2(1-p){\rm \ln}(1-p)-p)
\label{entrR}
\end{equation}
and $r$ is the lattice coordination number, from Eq. (\ref{dla}) one has
\begin{equation}
    \ln z = \ln(\frac{p}{r(1-p)^2}) -2\sum_{k=2}^{\infty} ka_kp^{k-1}
\label{logz}
\end{equation}
Substituting in Eq. (\ref{entr}) ${\rm ln} z$ from Eq. (\ref{logz}) and
$\lambda$ from Eqs.(\ref{entr2}, \ref{entrR}) one obtains
\begin{equation}
P = -{\rm \ln}(1-p) - \frac{p}{2} + \sum_{k=2}^{\infty}(1-k)a_k p^k
\label{eqP}
\end{equation}
so that a simple relation is obtained between the coefficients $m_k$ of the
virial expansion and the coefficients $a_k$ of the entropy expansion
\begin{equation}
    m_k = (k-1)(\frac{1}{k(k-1)} - a_k)
\label{mayer}
\end{equation}

In the case of hypercubic lattices in any dimension $d$ the 
coefficients $a_k$ have been computed in [\onlinecite{bfp}] 
through the order $20$; in that reference
in the cases $d=2,3,4$ there are the values through order $24$.
In appendix A we list the corresponding virial coefficients for $d=2,3$.

On these lattices, we can express the coefficients of
 the pressure, of the virial and the entropy expansions as simple
 polynomials in the variable $1/d$.
 Using the expressions for $a_k$ with $k \le 20$  in
 Ref.[\onlinecite{bfp}] to examine the real roots of $m_k$,
 reported in Table \ref{tabr}, and observing that, for large $d$, the leading
 coefficient of $m_k$  is positive, it follows that
 $m_k$ is positive for any integer $d$ with $d \ge 1$.

For example for $k=10$ one has
\begin{equation}
m_{10}=\frac{1024d^9 - 35712d^4 + 123240d^3 - 118260d^2 - 36990d + 67721}{10240d^9}
\nonumber
\end{equation}
with the three real root given in Table \ref{tabr}, dividing the real axis
in regions in which $m_{10}$ has signs $-,+,-,+$; the positive integer values
of $d$ occur in the positive regions.

\begin{table}[ht]
\caption{ Real roots of $m_k$ for $k \le 20$}
\begin{tabular}{|c|c|c|c|c|}
 \hline
 $k=2$& 0.25 &  && \\
 $k=3$& -0.354 & 0.354 && \\
$k=4$& -0.859 &  && \\
$k=5$& none &  && \\
$k=6$& -0.239 &  && \\
$k=7$& -2.032 & 0.848 &&\\
$k=8$& -1.796& -0.557& 0.859&\\
$k=9$& 1.044& 1.257&&\\
$k=10$& -0.655& 1.029& 1.313&\\
$k=11$& -3.404& 0.998&&\\
$k=12$& -3.241& -0.125& 0.997&\\
$k=13$& 1.000097& 1.725&&\\
$k=14$& 0.9994 &&&\\
$k=15$& -4.657& 0.999997 &&\\
$k=16$& -4.617& 1.000004& 1.801&\\
$k=17$& 1.000000085& 1.963 &&\\
$k=18$& 0.99999993 &&&\\
$k=19$& -5.852& 0.999999998& 2.005& 2.396\\
$k=20$& -5.879& 1.000000001& 1.993 &\\
 \hline
\end{tabular}
\label{tabr}
\end{table}

As remarked in the introduction, all the virial expansion coefficients so far
computed for the lattice dimer models are positive.
 This leads us to conjecture that they are all positive on all infinite
 regular lattices.

Let us remark that on any lattice with coordination number $r$ one has
\begin{equation}
a_k = \frac{r^{1-k}}{2 k(k-1)}
\label{akb}
\end{equation}
for $k$ less than the girth (i.e. the number of edges in a shortest
cycle) of the lattice graph; in particular this is true to all orders
in the case of the Bethe lattice [\onlinecite{nagle, stil}].

Note that from the assumption that the virial coefficients are all
positive, one gets an upper bound on the $a_k$,

\begin{equation}
    a_k \le \frac{1}{k(k-1)}
\label{quousquetandem}
\end{equation}

\section{An argument for  the bounds conjectured in Eq. (\ref{Delta0}) and Eq. (\ref{hlk}). }
In Ref.[\onlinecite{bfppos}], we introduced the Newton series for the
dimer entropy of a graph, in terms of the quantities
\begin{equation}
d(i)= {\rm \ln}(\frac {N(i)} {N(1)^i}) -{\rm \ln}(\frac {\bar N(i)} 
{\bar N(1)^i})
\label{di}
\end{equation}
\noindent
Here $N(i)$ is the number of configurations of $i$ dimers on the
graph $G$ with $v$ vertices and
 $\bar N(i)$ given by 
\begin{equation}
\bar N(i) = \frac{\bar{v}!}{(\bar{v}-2i)!i!2^i}
\label{nbar}
\end{equation}
 is the number of configurations of $i$ dimers on the complete graph on 
$\bar v\equiv 2\nu$ vertices, where $\nu$ is the matching number of $G$.
If the graph $G$ has  a perfect matching $v = \bar{v}$.

For a graph that satisfies the ``graph positivity'' property
introduced in Ref.[\onlinecite{bfppos}] one has

\begin{equation}
\Delta^k[d](i) \geq 0
\label{Delta}
\end{equation}
\noindent
where $k=0,...,\nu$ and $i=0,...,\nu-k$.
The validity of these bounds for all allowed $k$ and $i$ is in fact 
equivalent to their validity for all allowed $k$ and $i=0$. 

If the ``graph positivity'' {\it conjecture} of
Ref. [\onlinecite{bfppos}] is true, then eq.(\ref{Delta}) holds for
almost all regular biconnected bipartite graphs (in a reasonable
sense). More precisely, if this conjecture is true, then for each $r$
the fraction of $r$-regular biconnected bipartite graphs with $v$
vertices that satisfy eq.(\ref{Delta}) tends to $1$ as $v \to \infty$.
This positivity property is often satisfied also in non-regular
bipartite graphs, while it is usually not satisfied by non-bipartite
graphs.

We have a similar
situation here. Based upon the conjecture that all virial coefficients
are positive for dimer models on infinite regular lattices,
we state the following {\it conjecture}: 
the fraction of regular biconnected graphs with $v$ vertices that satisfy

\begin{equation}
\Delta^k[d](i) \leq  \Delta^k {\rm \ln}\frac{(\bar v -2i)!\bar v^{2i}} {\bar v !}
\label{Delta1}
\end{equation}
\noindent
tends to 1 as $v \to \infty$, when $k=2,...,\nu$ and $i=0,...,\nu-k$.  
The validity of these bounds for all allowed $k$ and $i$ is in fact 
equivalent to their validity for all allowed $k$ and $i=0$. 
Unfortunately, we do not know how to state mathematically 
how preponderantly this property holds for finite $v$. 
This will be seen ``experimentally'' in the following sections.
We now trace the path which leads from positivity of the virial 
series coefficients to Eq. (\ref{Delta1}) in the case of finite lattices.

With $p \approx \frac{2i}{v}$, in the limit of large $v$ one has
(see Eq. (11) in [\onlinecite{bfppos}])
\begin{equation}
\frac{1}{v}d(i) \to \sum_{k=2}^{\infty} a_k p^k
\label{da}
\end{equation}

We eliminate $a_k$ using Eq. (\ref{mayer}) and observe that
\begin{equation}
    \frac{1}{v}{\rm \ln}\frac{(v-2i)!v^{2i}}{v!} \to \sum_{k=2}^{\infty} \frac{p^k}{k(k-1)}
\end{equation}

Thus we get
\begin{equation}
\frac {1}{v}d(i)-\frac {1}{v}   {\rm \ln}\frac {(v-2i)!v^{2i}} {v!} \to 
- \sum_{k=2}^{\infty}\frac{1}{(k-1)}m_kp^k
\label{d1}
\end{equation}

Using $\frac{v}{2}\Delta \approx \frac{d}{d p}$ [\onlinecite{bfppos}],
assuming that the $m_k$ are positive and that $\frac{\bar v}{v} \to 1$
for $v \to \infty$, we are led to Eq. (\ref{Delta1}), in which we used
$\bar{v}$ instead of $v$, motivated by the fact that the tests of
Eq. (\ref{Delta1}) have fewer violations this way.  The assumption that
$\frac{\bar v}{v} \to 1$ is verified for biconnected $3$-regular
graphs and for regular bipartite graphs, since in these cases there
exists a perfect matching (see [\onlinecite{biedl}] and references
within). In fact all the biconnected regular graphs systematically
examined in next section have $\nu=n$, where $n = [\frac{v}{2}]$.

Using Eqs.(\ref{di}, \ref{nbar}) we can rewrite Eq. (\ref{Delta1}) as
\begin{equation}
    \Delta^k {\rm \ln} N(i) \le \Delta^k\left(i {\rm \,\ln}(\frac{N(1){\bar v}}{{\bar v}-1}) - {\rm \ln}(i!)\right)
\label{Delta1a}
\end{equation}
For $k \ge 2$, these bounds reduce to Eq. (\ref{Delta0}).

From the bound  Eq. (4) in [\onlinecite{HL1}], setting in that equation
$Z_i = N(n-i)$ and $M = n$, with $n \equiv [\frac{v}{2}]$, we get
\begin{equation}
  \Delta^2 {\rm \ln} N(i) \le {\rm \ln} \frac{(i+1)(n - i - 1)}{(i+2)(n - i)} = -\Delta^2 {\rm \ln}(i!(n-i)!).
\label{hl}
\end{equation}
It follows that, for $k=2$, the bounds of Eq. (\ref{Delta0}) are valid for any
graph.

Initially Eq. (\ref{Delta1}) would seem  to hold only for the graphs,
for example the periodic cubical graphs, whose limits are used to get the
given lattice functions. But we will try to apply Eq. (\ref{Delta1}),
or equivalently Eq. (\ref{Delta0}), to regular biconnected graphs
and to finite lattices.

\subsection{Rigorous upper bounds on the number of matchings}
The bound Eq. (\ref{hl}) follows from the fact that all
roots of the matching generating polynomial $M(z)= \sum_{i=0}^\nu N(i)z^i$
are negative[\onlinecite{HL}], so that the quantities
$P_\nu(i) = N(i)/\binom{\nu}{i}$
satisfy the Newton's inequalities $P(i+1)P(i-1) \le P(i)^2$
, or equivalently

\begin{equation}
\Delta^2 {\rm \ln}P(i) \le 0
\label{hl2a}
\end{equation}
for $i = 0,.., \nu-2$.
Using these inequalities and $\nu \le n$, where $n = [\frac{v}{2}]$,  it is easy to see that also
\begin{equation}
P(i) = \frac{N(i)}{\binom{n}{i}}
\label{Pi}
\end{equation}
satisfies them for $i = 0,.., n-2$.

Eq. (\ref{hl2a}) leads to the bounds (see [\onlinecite{frenkel}] and
Eq. (\ref{g4}) of Appendix C for the quantity $g(i) = {\rm
  \ln} P(i)$ with $k=2$)
\begin{equation}
P(i) \le \left(\frac{P(i_0+1)}{P(i_0)}\right)^{i-i_0} P(i_0)
\label{hl2e}
\end{equation}
for $i \ge i_0$.  In the case of general graphs, we can apply
Eq. (\ref{hl2e}) with $i_0=0$, using $N(0) = 1$ and $N(1) = E$, where
$E$ is the number of edges of the graph. 

In the following we will obtain a series of bounds based on Eq. (\ref{g4})
with given $k$ and $i_0$; we will denote such bounds as ``BX$k.i_0$'',
where B stands for Bound, X in G for a General graph, R for a Regular graph
and B for Bipartite regular graph; $k$ and $i_0$ are indices in Eq. (\ref{g4}).
The case $k=2$ corresponds to the Heilmann-Lieb inequality, holding for
all graphs. The case $k=3$, discussed in the next subsection, is conjectural;
to emphasize that we add the letter c to the name of the bound in the case $k=3$.

The choice of $i_0$ depends on the known $N(i)$ for a category of graphs;
for a general graph one has $N(0) = 1$ and $N(1) = E$.

Thus we get an upper Bound
for General graphs that we call BG2.0 to indicate that it follows from
Eq. (\ref{hlk}) with $k=2, i_0=0$,
\begin{equation}
N(i) \le (E/n)^i \binom{n}{i}
\label{hl2f}
\end{equation}
From this a simple bound for the matching generating polynomial follows

\begin{equation}
|M(z)| = |\sum_{i=0}^n N(i) z^i| \le \left(1 + \frac{E|z|}{n}\right)^n
\label{ho}
\end{equation}
Using $P_\nu(i)$ instead of $P(i)$ one gets Eq. (\ref{ho}) with $n$
replaced by $\nu$;
this stricter bound has been derived in [\onlinecite{carroll}].

For $E$ fixed, $v \le 2E$.
For the Hosoya index $Z = M(1)$,
introduced [\onlinecite{hosoya}] in theoretical chemistry to
characterize the topological structure of large molecules,
 when $v = 2E$ the bound Eq. (\ref{ho}) is saturated
by the graph $n K_2$. As $v \to \infty$, for $z= 1$ the bound 
Eq. (\ref{ho}) tends
to the bound $Z < e^E$, first derived in [\onlinecite{gutman}].

In the following we will derive other bounds for $P(i)$; one could
write down analogous bounds using $P_\nu(i)$, when $\nu < n$.

For a graph with $E$ edges and vertices of degrees $\delta_i$, we have
\begin{equation}
N(2) = \binom{E}{2} - \sum_{i=1}^v \binom{\delta_i}{2}
\end{equation}
 with
\begin{equation}
\sum_{i=1}^v \delta_i = 2E
\end{equation}

 Define $r = [\frac{2E}{v}]$ and $h = 2E - r v$, then one has $0 \le h < v$.
 Defining $\delta_i = r + k_i$, we get $\sum_{i=1}^v k_i = h$
 and 
\begin{equation}
\sum_{i=1}^v \delta_i^2 = v r^2 + 2r\sum_{i=1}^v k_i + \sum_{i=1}^v k_i^2
 \ge v r^2 + (2r + 1)\sum_{i=1}^v k_i
\end{equation}
  so that
\begin{equation}
\sum_{i=1}^v \delta_i^2 \ge vr^2 + (2r + 1)(2E-rv)
\label{dbnd}
\end{equation}

It follows that
using the bound Eq. (\ref{hl2e}) with $i_0=1$, we arrive at the  bound
 BG2.1, slightly tighter than BG2.0 for $i > 0$

\begin{equation}
N(i) \le \frac{E}{n}\left(\frac{2 N_m(2)}{(n-1)E}\right)^{i-1}
  \binom{n}{i}
\label{bnd2a}
\end{equation}
where $N_m(2)$ is the maximum value of $N(2)$
\begin{equation}
N_m(2) = \frac{E(E+1) - vr^2 - (2r+1)(2E-rv)}{2}
\label{n2}
\end{equation}
with
\begin{equation}
r \equiv \Big[ \frac{2E}{v} \Big]
\label{n2a}
\end{equation}

In the case of the BG2.1 bound, for the matching generating polynomial we obtain
\begin{equation}
|M(z)| \le 1 + 
\frac{(n-1)E^2}{2n N_m(2)}\left(\left(1 + \frac{2 |z| N_m(2)}{(n-1)E}\right)^n - 1\right)
\end{equation}
As an example, consider the
logarithm of the Hosoya index $Z$ for a general graph with $v=60$ and $E=110$.
Then the upper bound BG2.0 for this quantity has the value
$46.2$ while the bound BG2.1 has the value $45.6$.

In [\onlinecite{fried}] it has been shown that for a regular graph
with even $v$, the $N(i)$ satisfy the following bounds for $0 \le i \le n$;
the bound 
\begin{equation}
    N(i) \le 2^{-i}\binom{2n}{i} r^i
\label{bnd1R}
\end{equation}
which is tight in the region of low dimer density  and
\begin{equation}
N(i) \le \binom{2n}{2i}(r!)^\frac{i}{r}
\label{bnd2R}
\end{equation}
which is tight in the region of high dimer density.
In [\onlinecite{kahn}] another bound tight in the  region of high dimer 
density is given for $i < n$
\begin{equation}
N(i) \le exp \left(\frac{v}{2}(p {\rm \ln}(r) - p{\rm \ln}(p) -
2(1-p){\rm \ln}(1-p) - p + \frac{ {\rm \ln}(r)}{r-1} )\right)
\label{bnd2k}
\end{equation}
where $p = \frac{i}{n}$.
We shall denote by BR, the bound for the regular graphs
that  is the minimum among these three bounds.
For $r$ constant and $n$ large, the bound Eq. (\ref{hl2f}) improves
 Eq. (\ref{bnd1R}) by a factor slightly larger than\begin{equation}
\frac{(1-p)^{n-i+\frac{1}{2}}}{(1-\frac{p}{2})^{2n-i+\frac{1}{2}}}
\nonumber
\end{equation}
for $p$ not close to $0$ or $1$; for instance if $i=\frac{n}{2}$,
this factor is $1.09^n$.
Therefore a stricter
bound (that we denote as BR2.0), is obtained combining Eq. (\ref{hl2f})
and Eq. (\ref{bnd2R}).  A slightly stricter bound (that we call BR2.1)
is obtained combining Eq. (\ref{bnd2a}) and Eq. (\ref{bnd2R}).

In the case of regular bipartite graphs the following inequalities
are known [\onlinecite{fried}] for $0 \le i \le n$
\begin{equation}
N(i) \le \binom{n}{i} r^i
\label{bnd1B}
\end{equation}
giving a tight bound in the region of low dimer density  and
\begin{equation}
N(i) \le \binom{n}{i}^2 (r!)^\frac{i}{r}
\label{bnd2B}
\end{equation}
giving a tight bound in the region of high dimer density.  They lead
to the bound Eq. (3.3) in [\onlinecite{fried}] (that we denote as BB).
The bound Eq. (\ref{hl2f}) is the same as Eq. (\ref{bnd1B}) for these
graphs.  We can obtain a bound (denoted as BB2.2), which is stricter
than BB, by using Eq. (\ref{hl2e}) with $i_0=2$, and observing that
\begin{equation}
N(2) = \frac{nr(nr - 2r + 1)}{2}
\label{n2reg}
\end{equation}
and [\onlinecite{friedbnd}]
\begin{equation}
N(3) = \binom{nr}{3} - 2n\binom{r}{3} - nr(r-1)^2 - 
2n\binom{r}{2}(nr - 3r + 2)
\label{n3}
\end{equation}

Using Eq. (\ref{hl2e}) with $i_0=3$, using $N(3)$ and the maximum value
of $N(4)$, given by [\onlinecite{friedbnd}]
\begin{equation}
N_m(4) = \frac{n^4r^4}{24} + \frac{n^3r^3}{4}(1-2r) +
\frac{n^2r^2}{24}(19-60r+52r^2) + 
nr(\frac{5}{4} - 5r + 7r^2 - \frac{7r^2}{2}) + \frac{rn(r-1)^2}{4}
\label{n4}
\end{equation}
we get a bound slightly better than choosing $i_0 = 2$.  For $r$ fixed,
$v$ large, and $i_0 = 3$, the bound Eq. (\ref{hl2e}) for $N(i)$ is
smaller than the bound Eq. (\ref{bnd1B}) by roughly a factor 
$\exp({(3 -\frac{3}{r})\frac{i-2}{n}})$. 
For instance with $i = \frac {n}{2}$ and $n$
large, the bound is smaller by a constant factor
$\exp({\frac {3}{2}(1-\frac {1}{r})})$.
We denote by BB2.3 the bound Eq. (\ref{hl2e}) for $i_0 = 3$ combined with
the bound Eq. (\ref{bnd2B}).

\subsection{Conjecture of a stricter bound  on virial coefficients}

We now set $p = p(i) = 2i/v$ and assume that $v$, $n-i$ are large.
From Eq. (\ref{d1},\ref{nbar},\ref{di}) it follows that for $k \ge 2$
\begin{equation}
\Delta^k(\frac{1}{v} {\rm \ln}(N(i) i!) \approx
-\Delta^k \sum_{j=2}^{\infty} \frac{m_jp^j}{j-1}
\label{b2}
\end{equation}
from which it follows that
\begin{equation}
    \Delta^k(\frac{1}{v} {\rm \ln}(N(i) i!(n-i)!) \approx 
\Delta^k \sum_{j=2}^{\infty}(\frac{1}{2j} - m_j)\frac{p^j}{j-1}
\label{Delta3}
\end{equation}
All the known virial coefficients in the infinite regular lattice
models mentioned in the introduction satisfy the bound
\begin{equation}
m_k \ge \frac{1}{2k}
\label{m2}
\end{equation}
for $k \ge 2$.
We have also checked that, in $d$ dimensional hypercubic infinite
lattices the virial expansion coefficients satisfy the bound
Eq. (\ref{m2}) through order $20$ for any $d$, so that from Eq. (\ref{Delta3})
it follows that Eq. (\ref{hlk}) is satisfied in these cases.
We are therefore lead to conjecture that Eq. (\ref{m2}) holds on all
infinite regular lattices. This indicates that there is a singularity
in the virial expansion of the pressure, for $p=1$.
As already indicated in [\onlinecite{bfp}] the entropy is analytic 
in the interval $0 \le p < 1$ and it is bounded above at $p=1$.
However from Eq. (\ref{dla}) if follows that $\frac{d \lambda}{d p}$
is log-singular for large $z$ (equivalently for $p \to 1$).
As to the pressure, from Eq. (\ref{entr}) and the fact that
$\lambda(p)$ is bounded above at $p=1$ and $p(z) \to 1$ as 
$z \to\infty$, it follows that the pressure is log-singular for 
large $z$.

Eq. (\ref{hlk}) for $k=2$ is satisfied for any graph, due to the
Heilmann-Lieb inequality Eq. (\ref{hl}).
We are therefore led to the conjecture that in all infinite regular
lattice models the virial coefficients satisfy Eq. (\ref{m2}), and to
the conjecture that the frequency of the violations to the bounds
Eq. (\ref{hlk}) for large regular biconnected graphs, tends to zero as
$v \to \infty$.

In [\onlinecite{FKM}], [\onlinecite{FKM1}] it has been conjectured that 
in the limit of infinitely large
random bipartite regular graphs, the entropy has almost surely
the coefficients $a_k$ given in Eq. (\ref{akb}); the same is true for
the average distribution Eq. (\ref{Nfkm}) introduced in [\onlinecite{FKM}].
In the appendix we show that this average distribution satisfies the
bounds Eq. (\ref{hlk}), so that  this conjecture suggests that
the bipartite regular graphs almost surely satisfy Eq. (\ref{hlk}).

\begin{figure}[tbp]
\begin{center}
\leavevmode
\includegraphics[width=5.5 in]{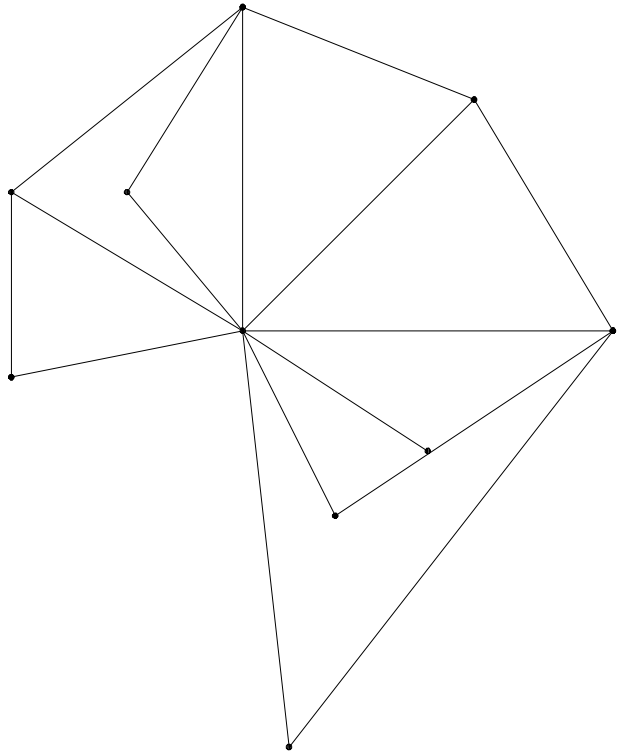}
\caption{ \label{fig1} A graph violating the bound Eq. (\ref{hlk}) for $k=3$.}
\end{center}
\end{figure}

In the tests discussed in the next section, no violation of the bound
Eq. (\ref{hlk}) with $k=3$ are observed both for regular biconnected graphs 
and for not necessarily regular lattice graphs.  Notice however that for
$k=3$, in the case of more general non-regular graphs there are
violations, e.g. in the case of the graph with perfect matching shown
in Figure \ref{fig1}.

It seems therefore interesting to investigate the bound Eq. (\ref{hlk})
for $k=3$ in the case of biconnected regular graphs.

Using Eq. (\ref{g4}) for $k=3$ we get
\begin{equation}
P(i) \le P(i_0)\left(\frac{P(i_0+1)}{P(i_0)}\right)^{i-i_0}
\left( \frac{P(i_0+2)P(i_0)}{P(i_0+1)^2}
\right)^{ \frac{(i-i_0)(i-i_0-1)}{2}  }
\label{fbnd}
\end{equation}
for $i \ge i_0$.

In the case of regular graphs, we can use Eq. (\ref{fbnd}) with
$i_0=0$, taking into account Eq. (\ref{n2reg}) for $N(2)$.  The
{\it conjectured}  bound BR3.0c is obtained combining this bound with Eq. (\ref{bnd2R}).

As an example of a regular graph, consider the Buckminster fullerene $C_{60}$;
it has $v=60$ and $E=90$; the first violation of Eq. (\ref{hlk}) is for $k=20$.
The logarithm of the Hosoya index $Z$ of this graph  has the value 
${\rm \ln}(Z) = 34.89$. The bound BR has the value $46.49$,
the bound BR2.0 yields $41.50$, the bound BR2.1 is $41.02$, while
BR3.0c is $36.58$.

In the case of regular bipartite graphs, we can use Eq. (\ref{fbnd})
with $i_0=2$, using $N(2), N(3)$ and replacing  $N(4)$ by $N_m(4)$,   
as in subsection IIIA.
We thus {\it conjecture} a bound (called BB3.2c)  obtained combining 
this bound with Eq. (\ref{bnd2B}).

When $r$ divides $n$, the bound BB3.2c is weaker than the 
Upper Matching conjecture (UMC)[\onlinecite{FKM}],
according to which the number of $i$-matchings of a $r$-regular bipartite 
graph with
$n=q r$ is  bounded above by the number of $i$-matchings of $q K_{r,r}$,
for $q$ a positive integer.

In Table \ref{boundsBB} we give ${\rm \ln} Z$, where $Z$ is the Hosoya index,
and upper bound estimates for two graphs, the periodic $12\times12$ square
grid and the $6$-cube; for the latter we took the matching polynomial
from [\onlinecite{oeis}] which used [\onlinecite{Lundow1, Lundow2}].

\begin{table}[ht]
  \caption{${\rm \ln} Z$, where $Z$ is the Hosoya index and upper bounds
    estimates for two bipartite regular graphs. 
    The first graph is the periodic grid $12\times12$, with $V=144$ and degree
    $4$; the second one is the $6$-cube, with $V=64$ and degree $6$.
    The latter two upper bounds BB3.2c and UMC are conjectural. 
    The UMC conjecture  applies only to graphs
    in which  $V/2$ is a multiple of the degree,  so that for the $6$-cube  
    no entry appears in the Table.
  }
\begin{tabular}{|c|c|c|c|c|c|}
 \hline
 graph                   &${\rm \ln} Z$& BB  &BB2.3    & BB3.2c & UMC\\
 \hline
periodic grid $12\times12$& $95.44$&  $115.24$& $113.69$&  $100.11$& $96.16$\\
$6$-cube                 & $50.32$&  $59.59$ & $58.49$ & $52.87$&          \\
 \hline
 \end{tabular}
 \label{boundsBB}
\end{table}

The $d$-cubes through $d=6$ satisfy the bounds Eq. (\ref{Delta0}) and
Eq. (\ref{hlk}).

\section{Tests on the upper bounds}
We shall now review our graph tests.  All graphs we considered are
simple biconnected graphs.  The tests of Eqs.(\ref{Delta0}) and
Eqs.(\ref{hlk}) are done for all allowed values of $i$ (one could
check only the case $i=0$ to see if there are violations for some $k$,
but the value of $k$ for which there is a violation is generally
higher if one considers only the case $i=0$).

We made tests on lattice graphs and on regular graphs; to generate
systematically the latter we have used the {\it geng} program in the {
  \it Nauty} package[\onlinecite{nauty}], via the {\it Sage}
interface[\onlinecite{Sage}].  The matching generating polynomials are
computed with the aid of the algorithm described in
[\onlinecite{bpalg}].  To perform our computations, we have used an
ordinary desktop personal computer based on a processor Intel $i7$
$860$ with a RAM of 8 $GB$.

\subsection{Tests of the bounds Eq. (\ref{Delta0}) 
on finite lattice graphs. i) Periodic boundary conditions}
Let us first discuss the tests on finite lattices with periodic $bc$.
In the case of rectangular grids of size $L_x\times L_y$ with $L_x \ge
L_y$, we considered the cases listed in
Tab. \ref{tabdn1}:

\begin{table}[H]
  \caption{Violations of the bounds Eq. (\ref{Delta0}) 
 in the case of rectangular grids 
 of size $L_x\times L_y$ , with $L_x   \ge L_y$ and periodic $bc$.
  }
\begin{tabular}{|c|c|c|}
\hline
  $L_x$ & $L_y$ & Violations of the bounds Eq. (\ref{hlk})\\
\hline
$L_x \le 2000$ & $ L_y=3$ &  for  $L_x \ge 9$\\
$L_x \le 1100$ & $ L_y=4$ &  for all $L_x \ge 409$\\
$L_x \le 500$ & $ L_y=5$&   for $L_x \ge 186$\\
 $L_x \le 200$ & $  L_y=6$  & none\\ 
 $L_x \le 70$ & $  L_y=7$  & none\\  
 $L_x \le 50$ & $  L_y=8$  & none\\ 
 $L_x \le 20$ & $  L_y=9$  & none\\ 
 $L_x \le 15$ & $  L_y=10$  & none\\ 
 $L_x = 11$ & $  L_y=11$  & none\\ 
 $L_x = 12$ & $  L_y=12$  & none\\ 
 \hline
 \end{tabular}
 \label{tabdn1}
\end{table}

Therefore we have found violations only for $L_y \le 5$, when $L_x \ge 3L_y $. 
The minimum $k$ for which there are violations
is $12$ for the cases with $L_y=3$, it is larger than $100$ in the other cases.

In the case of triangular grids of size $L_x\times L_y$ with $L_x \ge
L_y$, we have considered the cases listed in
Tab. \ref{tabdn2}.

\begin{table}[H]
  \caption{Violations of the bounds Eq. (\ref{Delta0})  in the case 
    of triangular grids 
    of size $L_x\times L_y$ , with $L_x   \ge L_y$ and periodic $bc$.
    These grids are obtained by
    adding a SW-NE diagonal to rectangular grids $L_x \times L_y$.
  }
\begin{tabular}{|c|c|c|}
 \hline
  $L_x$ & $L_y$ & Violations of the bounds Eq. (\ref{hlk})\\
\hline
$L_x \le 1000$ & $ L_y=3$ & for all $L_x \ge 16$\\
$L_x \le 1000$ & $ L_y=4$ &  none\\
$L_x \le 400$ & $ L_y=5$&   for $L_x \ge 66$  \\
 $L_x \le 200$ & $  L_y=6$  & none\\ 
 $L_x \le 50$ & $  L_y=7$  & none\\ 
 $L_x \le 17$ & $  L_y=8$  & none\\ 
 \hline
 \end{tabular}
 \label{tabdn2}
\end{table}

In the case of hexagonal lattices of size $L_x \times L_y$ with
periodic $bc$ (in the brick-wall representation) with $L_x, L_y$ even,
we have considered the cases listed in Tab. \ref{tabdm3}.
\begin{table}[H]
 \caption{Violations of the bounds Eq. (\ref{Delta0})  in the case 
    of hexagonal grids 
    of size $L_x\times L_y$ , with $L_x   \ge L_y$ and periodic $bc$.
  }
\begin{tabular}{|c|c|c|}
 \hline
 $L_x$ & $L_y$ & Violations of the bounds Eq. (\ref{hlk})\\
\hline
$L_x=4$ & $  4 \le L_y \le 100$& all cases\\
$L_x=6$ & $  4 \le L_y \le 100$& for $L_y \ge 26$\\
 $8 \le L_x \le 14 $ & $ 4 \le L_y \le 14$& none \\
$6 \le L_x \le 100 $ & $ L_y=4$& none\\
 $6 \le L_x \le 100 $ & $ L_y=6$& none\\
\hline
 \end{tabular}
 \label{tabdm3}
\end{table}

The  minimum value of $k$ for which we found a violation is $7$.
In particular there are no violations for $L_x$ close to $L_y$ and $L_x > 4$.

\subsection{Tests of the bounds Eq. (\ref{Delta0}) 
on finite lattice graphs. ii) Open boundary conditions}
Let us now  turn to finite lattices with open boundary conditions.

In the case of rectangular grids of size $L_x \times L_y$, with open
$bc$, we have examined the cases $2 \le L_y \le L_x \le 19$, finding
no violations to Eq. (\ref{Delta0}).

In the case of triangular grids of size $L_x \times L_y$, obtained by
adding a SW-NE diagonal in a rectangular grids $L_x \times L_y$ with
open $bc$, for $2 \le L_y \le L_x \le 18$ there are violations only
for $L_y \times 3$ with $L_x \ge 10$.  The minimum value for which
Eq. (\ref{Delta0}) is violated is $k=16$.

We have also examined three-dimensional slabs of size $L_x \times L_y
\times L_z$ with open $bc$. The results are summarized in
Tab. \ref{tabdn4}.
\begin{table}[H]
 \caption{Violations of the bounds Eq. (\ref{Delta0})  in the case 
    of three-dimensional slabs of size $L_x \times L_y \times L_z$
    with  open $bc$.
  }
\begin{tabular}{|c|c|c|c|}
 \hline
 $L_x$ & $L_y$ & $L_z$ & Violations of the bounds Eq. (\ref{hlk})\\
\hline
$2 \le L_x \le 1000$ & $L_y=3$ & $L_z=2$ &none\\
 $3 \le L_x \le 500$ & $  L_y=3$ & $  L_z=3$ &none\\
 $4 \le L_x \le 600$ & $  L_y=4 $ & $ L_z=2$ &none\\
 $4 \le L_x \le 200$ & $  L_y=4 $ & $ L_z=3$&none\\
 $4 \le L_x \le 40 $ & $ L_y=4$ & $  L_z=4$ &none\\
\hline
 \end{tabular}
 \label{tabdn4}
\end{table}

For all the graphs examined in this section, Eq. (\ref{Delta0}) is
satisfied for $k \le 4$. It would be interesting to know whether these
bounds, Eq. (\ref{Delta0}) for $k \le 4$, are always satisfied for
regular biconnected graphs.

\subsection{Tests of the bounds Eq. (\ref{Delta0}) on bipartite graphs}
We have tested the validity of the upper bounds Eq. (\ref{Delta1})
(equivalently Eq. (\ref{Delta0})) for regular bipartite biconnected
graphs (RBB), by enumerating and studying exhaustively a large class
of graphs.  For $v \le 18$ vertices, we observe a single violation in
the case of a $3$-regular graph with $v=18$.  For $3$-regular graphs
with $ 18 \le v \le 30$, the frequency of the violations decreases
with increasing $v$ ($v$ is even, since there are no graphs with $v$
odd in this class).

It is interesting to observe that, in the cases considered in
Table \ref{tabdeg3}, the average order of the automorphism groups of
the positivity-violating graphs is a few times larger
than the average order of the automorphism groups of all the
RBB graphs with the same vertex degree. The same is true in the following
tests of this section.

\begin{table}[ht]
  \caption{For the RBB graphs with a given number  $18 \le v \le 30$ of 
    vertices of degree 3, we have listed the number of graphs in this
    class, the number of violations of the upper bounds, the average
    order $ng$ of the automorphism groups of  the graphs, the average order
    $ngv$ of this group for the graphs violating the bounds
    Eq. (\ref{Delta0}).  $k$ is the minimum value for which
    Eq. (\ref{Delta0}) is violated.  }
\begin{tabular}{|c|c|c|c|c|c|}
 \hline
 $v$ & number of graphs & violations & $ng$ & $ngv$ & $k$\\
 \hline
18&   149&   1&  15.1 & 64. & 9\\
20&   703&   3&  8.7 &  91. & 10\\
22&   4132&  13& 4.5 &  40. & 9\\
24&   29579& 38& 3.3 &  32. & 7\\
26&   245627& 253& 2.3 & 22. & 7\\
28&   2291589& 1392& 1.9&  20. & 6\\
30&   23466857& 8008& 1.7& 16. &7\\
 \hline
 \end{tabular}
 \label{tabdeg3}
\end{table}

For the RBB graphs with vertices of degree $4$ there is one violation
for $v=20$ among $62611$ graphs and $k=10$ is the minimum value for
which Eq. (\ref{Delta0}) is violated.  The order of the automorphism
group of the violating graph is $256$, while the average order is
$3.1$.  For $v=22$, there are $5$ violations among $2806490$ graphs
and the minimum value for which Eq. (\ref{Delta0}) is violated is
$k=11$; the average of the orders of the automorphism groups of the
violating graphs is $3721.6$, while the average order of all graphs
is $1.5$.

For the RBB graphs with vertices of degree $5$, there is one violation
for $v=20$ out of $304495$ graphs, with $k=9$; the order of the
automorphism group of this graph is $1327104$, while the average order
is $7.1$.

For the RBB graphs with vertices of degree larger than $4$, we had to
restrict to graphs with $v \le 20$ vertices and observed no violations
for degree larger than $5$.

\subsection{Tests of the bounds Eq. (\ref{Delta0}) on non-bipartite graphs}

In the case of the biconnected $3$-regular non-bipartite graphs with
 $v \le 22$, the first violation occurs for $v=12$.
For $v \ge 12$ the frequency of the violations decreases with $v$,
as shown in Table \ref{tabdeg3n}.

\begin{table}[ht]
 \caption{For the biconnected $3$-regular non-bipartite graphs with a
 given number $v$ of vertices, we have listed the number of
 graphs in this class, the number of violations of the upper bounds,
 the average order $ng$ of the automorphism groups of   the graphs, the
 average order $ngv$ of this group for the graphs violating the bounds
 Eq. (\ref{Delta0}).  The cases with $v$ odd are not listed, since
 there are no such graphs.  $k$ is the minimum value for which
 Eq. (\ref{Delta0}) is violated.  }
\begin{tabular}{|c|c|c|c|c|c|}
 \hline
 $v$ & number of graphs & violations & $ng$ & $ngv$ & $k$\\
 \hline
12&   76&   1&  7.4& 16. & 6\\
14&   467&  6&  4.4&  8.7 & 6\\
16&   3836& 44& 3.1&  8.3 & 5\\
18&   39717& 257& 2.2& 7.5 & 5\\
20&   497115& 2856& 1.7& 5.6 & 5\\
 22&   7183495& 29597& 1.5& 4. & 5\\
 \hline
 \end{tabular}
 \label{tabdeg3n}
\end{table}

In the case of the biconnected $4$-regular non-bipartite graphs with
$v \le 17$, the first violations occur for $v=12$, as shown
in Table \ref{tabdeg4n}.

\begin{table}[ht]
 \caption{For the biconnected $4$-regular non-bipartite graphs with a
 given number $v$ of vertices , we have listed the number of
 graphs in this class, the number of violations of the upper bounds,
 the average order $ng$ of the automorphism groups of   the graphs, the
 average order $ngv$ of this group for the graphs violating the bounds
 Eq. (\ref{Delta0}).  $k$ is the minimum value for which
 Eq. (\ref{Delta0}) is violated.  }
\begin{tabular}{|c|c|c|c|c|c|}
 \hline
 $v$ & number of graphs & violations & $ng$ & $ngv$ & $k$\\
 \hline
12& 1538&   2&  3.4& 40 & 6\\
13& 10768&  0&     &    &\\
14& 88112&  12& 1.6& 17 & 6\\
15& 805281& 30& 1.3& 14.& 7\\
16& 8036122& 454& 1.2& 14. & 6\\
17& 86214189& 295& 1.2& 10. & 6\\
 \hline
 \end{tabular}
 \label{tabdeg4n}
\end{table}

Unlike in the bipartite case and in the case of $3$-regular
non-bipartite graphs examined above, there exist graphs with odd $v$.

 The frequency of the violations occurring among the graphs with even
 $v$ decreases regularly as $v$ increases. The same is true for the
 graphs with odd $v$.

\subsection{Tests of the bounds Eq. (\ref{hlk}) on finite lattice graphs.
 i) Periodic boundary conditions}
Let us first discuss the tests on finite lattices with periodic $bc$.

In the case of rectangular grids of size $L_x\times L_y$ with $L_x \ge
L_y$, we have considered the cases listed in
Tab. \ref{tabdnn1}:

\begin{table}[H]
  \caption{Violations of the bounds Eq. (\ref{hlk}) 
  in the case of rectangular grids 
    of size $L_x\times L_y$ , with $L_x   \ge L_y$ and periodic $bc$.
  }
\begin{tabular}{|c|c|c|}
 \hline
  $L_x$ & $L_y$ & Violations of the bounds Eq. (\ref{hlk})\\
 \hline
$L_x \le 2500$& $L_y=3$ & for all $L_x \ge 5$\\
$L_x \le 1000$& $L_y=4$ & for most $L_x \ge 8$\\
$L_x \le 700$& $L_y=5$ & for most $L_x \ge 9$\\
$L_x \le 200$& $ L_y=6$ & for most $L_x \ge 12$\\
$L_x \le 70$& $ L_y=7$ & none\\
$L_x \le 50$ & $L_y=8$ &  for most $L_x \ge 14$\\
$L_x \le 20$ & $ L_y=9$ &  none\\
$L_x \le 15$ & $L_y=10$ &  none\\
$L_x =11$ & $L_y=11$ &  none\\
$L_x =12$ & $L_y=12$ &  none\\
 \hline
 \end{tabular}
 \label{tabdnn1}
\end{table}
The minimum value of $k$ for which  violations are observed, is $4$.

In the case of the triangular grids of size $L_x\times L_y$ with $ L_x
\ge L_y$ and periodic $bc$, we considered the cases listed in
Tab.\ref{tabdnn2}.
\begin{table}[H]
  \caption{Violations of the bounds Eq. (\ref{hlk}) 
in the case of triangular grids 
    of size $L_x\times L_y$ , with $L_x   \ge L_y$ and periodic $bc$.  }
\begin{tabular}{|c|c|c|}
 \hline
  $L_x$ & $L_y$ & Violations of the bounds Eq. (\ref{hlk})\\
 \hline
$L_x \le 2500$ & $L_y=3$ & for most $L_x \ge 11$\\
$L_x \le 1000$ & $ L_y=4$ &for all $L_x \ge 88$\\
$L_x \le 500$ & $  L_y=5$ &for most $L_x \ge 29$\\
 $L_x \le 150$ & $ L_y=6$& none\\
$L_x \le 50$ & $  L_y=7$& none\\
$L_x \le 15$ & $ L_y=8$ &none\\ 
 \hline
 \end{tabular}
 \label{tabdnn2}
\end{table}
The minimum value of $k$ for which there are violations is $4$.

In the case of the hexagonal grids (in the brick-wall representation)
of size $L_x\times L_y$, with $L_x \ge L_y$ and periodic $bc$ we
considered the cases listed in Tab.\ref{tabdn3}.
\begin{table}[H]

  \caption{Hexagonal grids (in the brick-wall representation)
    of size $L_x\times L_y$, with $L_x   \ge L_y$ and periodic $bc$ tested .
  }
\begin{tabular}{|c|c|c|}
 \hline
 $L_x$ & $L_y$ & violations of the bound Eq. (\ref{hlk}) \\
 \hline
$L_x=4$ &  $4 \le L_y \le 100$& $L_y=6$ and $L_y >= 10$  \\ 

$L_x=6$ &  $4 \le L_y \le 100$& $L_y \ge 6$\\ 

$8 \le L_x \le 14$  &  $4 \le L_y \le 14$& none for $L_x - 8 \le L_y < L_x$ \\

$8 \le L_x \le 100$ & $L_y=4$ & $L_y \ge 14$ \\

$8 \le L_x \le 30$ & $L_y=6$ & $L_x \ge 18$  \\
 \hline
 \end{tabular}
 \label{tabdn3}
\end{table}
Summarizing this table, we found no violations 
in the band $max(4, L_x - 8) \le L_y < L_x$; almost
all the other cases violate the bounds.
The violations of the bounds Eq. (\ref{hlk}) occur for $k \ge 4$.

\subsection{Tests of the bounds Eq. (\ref{hlk}) on finite lattice graphs.
 ii) Open boundary conditions}

Let us now turn to finite lattices with open $bc$.

In the case of the rectangular grids of size $L_x\times L_y$ with $2
\le L_y \le L_x \le 19$ and open $bc$, there are many violations, but
they do not occur for $L_x=L_y$ with $L_x \ne 5$, or for $L_y \ge 9$.
The minimum value of $k$ for which Eq. (\ref{hlk}) is violated is
$k=4$.

In the case of the triangular grids $L_x\times L_y$ with $2 \le L_y
\le L_x \le 18$ and open $bc$, there are no violations for $L_x=L_y$
with $L_x \ne 5$, or for $L_y \ge 8$.  The minimum $k$ for which there
are violations to Eq. (\ref{hlk}) is $k=6$.

In the case of the 3-d grids of size $L_x \times L_y \times L_z$ with
open $bc$, we considered the  cases listed in Tab. \ref{tabdn5}.

\begin{table}[H]
 \caption{Violations of the bounds Eq. (\ref{hlk})  in the case 
    of three-dimensional slabs of size $L_x \times L_y \times L_z$
    with  open $bc$.}
\begin{tabular}{|c|c|c|c|}
 \hline
 $L_x$ & $L_y$ & $L_z$ & Violations of the bounds Eq. (\ref{hlk})\\
\hline
$L_x \le 1000$ & $ L_y=3$ & $ L_z=2$ & for $L_x \ge 8$\\
$L_x \le 500$ & $ L_y=3$ & $ L_z=3$ &   for $L_x=5, 9, 11, 13, 15, 17, 21$\\
$L_x \le 600$ & $ L_y=4$ & $ L_z=2$ &  for $L_x \ge 11$\\
$L_x \le 200$ & $ L_y=4$ & $ L_z=3$ &  for $L_x \ge 16$\\
$L_x \le 40$ & $ L_y=4$ & $ L_z=4$ & for  $L_x \ge 20$\\
\hline
 \end{tabular}
 \label{tabdn5}
\end{table}

In particular,  no violations are observed for $L_x= L_y= L_z$.

Summarizing, in the tests of Eq. (\ref{hlk}) on lattice graphs,
violations are observed more often than for Eq. (\ref{Delta0}); for
bidimensional lattices with $L_x \approx L_y >5 $ there are no
violations, except in the case of the hexagonal lattice, in which no
violations are present on a strip with $L_x > L_y$. This restriction
might be due to the fact that the hexagonal lattice is sensitive to
the boundary conditions[\onlinecite{elser}].  Therefore these tests
give some indication, although not as sharply as in the case of
Eq. (\ref{Delta0}), that in the limit of infinite lattices the bounds
Eq. (\ref{hlk}) and the corresponding bounds Eq. (\ref{m2}) are
satisfied, and virial positivity holds.

\subsection{Tests of the bounds Eq. (\ref{hlk}) on bipartite graphs}

Let us now discuss the systematic tests on regular graphs.

In the case of the RBB graphs with vertices of degree $3$ and $v \le
30$, the first violation occurs for $v=14$; for larger values of $v$
the frequency of violations decreases irregularly (e.g. it increases
at $v=20$), as shown in Table \ref{tabdeg3e}.

\begin{table}[ht]
 \caption{For the RBB graphs with a given number $14 \le v \le 30$ of
 vertices of degree 3, we have listed the number of graphs in this
 class, the number of violations of the upper bounds Eq. (\ref{hlk}),
 the average order $ng$ of the automorphism groups of   the graphs, the
 average order $ngv$ of this group for the graphs violating the bounds
 Eq. (\ref{hlk}); $k$ is the minimum value for which these bounds are
 violated.  }
\begin{tabular}{|c|c|c|c|c|c|}
 \hline
 $v$ & number of graphs & violations & $ng$ & $ngv$ & $k$\\
 \hline
14 &  13  & 1 & 44. & 28. & 7\\
16 &  38  & 2 & 19. & 48. & 6\\
18 &  149 & 5 & 15. & 84. & 5\\
20 &  703 & 33 & 8.7 & 29. & 5 \\
22 &  4132 & 106 & 4.5 & 20. & 4\\
24 &  29579 & 619 & 3.3 & 16. & 4\\
26 &  245627 &3415 & 2.3 & 10. & 4\\
28 &  2291589&22913& 1.9 &7.9  & 4\\
30 &  23466857&163789&1.7&6.2  & 4\\
 \hline
 \end{tabular}
 \label{tabdeg3e}
\end{table}

In the case of the RBB graphs with vertices of degree $4$ and $v \le
22$, the first violation occurs for $v=16$. For larger values of $v $
the frequency of the violations decreases, as shown in the Table
\ref{tabdeg4e}.

In the case of RBB graphs with vertices of degree $5$ and $v \le 20$,
the first $3$ violations occur for $v=20$ out of 304495 graphs, with
average order $449280$ of the automorphism groups of the violating
graphs, to be compared with an average order of $7.1$ for all the
graphs; the minimum $k$ for which there are violations is $k=6$.

We have checked that for RBB graph with  $v \le 20$, there are no 
violations for $v > 5$.

\begin{table}[ht]
 \caption{For the RBB graphs with $16 \le v \le 22$ vertices of degree 4,
 we have listed
 the number of graphs in this class, the number of violations of the
 upper bounds Eq. (\ref{hlk}), 
 the average order $ng$ of the automorphism groups of   the 
 graphs, the average order $ngv$ of this group for the graphs
 violating the bounds Eq. (\ref{hlk});
 $k$ is the minimum value for which these bounds are violated.
}
\begin{tabular}{|c|c|c|c|c|c|}
 \hline
 $v$ & number of graphs & violations & $ng$ & $ngv$ & $k$\\
 \hline
16 &  129  & 1  & 112. & 5184. & 8\\
18 &  1980  & 1 & 8.7  & 576.  & 7\\
20 &  62611 & 18& 3.1  & 1901. & 6\\
22 &  2806490 & 115 & 1.5 & 487. & 6\\
 \hline
 \end{tabular}
 \label{tabdeg4e}
\end{table}

\subsection{Tests of the bounds Eq. (\ref{hlk}) on non-bipartite graphs}

In the case of biconnected $3$-regular non-bipartite graphs with $v
\le 20$, the first violation occurs for $v=10$. For larger values of
$v$ the frequency of violations decreases irregularly as shown in
Table \ref{tabdeg3ne}.

\begin{table}[ht]
 \caption{For the biconnected non-bipartite graphs with a given number
$10 \le v \le 22$ of vertices of degree 3, we have listed the number
of graphs in this class, the number of violations of the upper bounds
Eq. (\ref{hlk}), the average order $ng$ of the automorphism groups of the
graphs, the average order $ngv$ of this group for the graphs violating
the bounds Eq. (\ref{hlk}); $k$ is the minimum value for which these
bounds are violated.  }
\begin{tabular}{|c|c|c|c|c|c|}
 \hline
 $v$ & number of graphs & violations & $ng$ & $ngv$ & $k$\\
 \hline
10 &  16  & 1  & 15. & 16. & 5\\
12 &  76  & 2  & 7.4 & 12. & 5\\
14 &  467 & 11 & 4.4 & 11. & 4\\
16 &  3836 & 102 & 3.1 & 9. & 4\\
18 &  39717 & 741 & 2.2 & 6.2 & 4\\
20 &  497115 & 7324 & 1.7 & 4.8 & 4\\
22 &  7183495& 78426&1.5  & 3.5 & 4\\
 \hline
 \end{tabular}
 \label{tabdeg3ne}
\end{table}

In the case of biconnected $4$-regular non-bipartite graphs
with $v \le 17$,  a first violation is met for $v=11$. For larger $v$, 
the frequency of the violations decreases regularly for $v$ even, while
it decreases irregularly for $v$ odd (it increases for $v=15$), 
see Table \ref{tabdeg4ne}.

\begin{table}[ht]
 \caption{For the biconnected non-bipartite graphs with 
     $10 \le v \le 17$ vertices of degree 4,
 we have listed
 the number of graphs in this class, the number of violations of the
 upper bounds Eq. (\ref{hlk}), 
 the average order $ng$ of the automorphism groups of the 
 graphs, the average order $ngv$ of this group for the graphs
 violating the bounds Eq. (\ref{hlk});
 $k$ is the minimum value for which these bounds are violated.
}
\begin{tabular}{|c|c|c|c|c|c|}
 \hline
 $v$ & number of graphs & violations & $ng$ & $ngv$ & $k$\\
 \hline
11 &  264  & 1  & 5.2 & 12. & 5\\
12 &  1538  & 3 & 3.4 & 32. & 5\\
13 &  10768 & 37 & 2.1 & 16. & 5\\
14 &  88112 & 34 & 1.6 & 31. & 5\\
15 &  805281 & 3086 & 1.3 & 5.0 & 4\\
16 &  8036122& 1121 & 1.2 & 14. & 5\\
17 & 86214189&197431& 1.2 & 2.5 & 4\\
 \hline
 \end{tabular}
 \label{tabdeg4ne}
\end{table}

In addition to the systematic examination of $3$-regular biconnected
graphs up to $v=22$, we have tested the $k=3$ bound for $v=30$.  Using
the {\it NetworkX} [\onlinecite{networkx}] random regular graph
generator, we have thus examined over $100$ millions $3$-regular graphs
with $v=30$, and have checked that Eq. (\ref{hlk}) is satisfied for
$k=3$.  From Table \ref{tabdeg3ne}, by a simple extrapolation we
estimate that for $v=30$ there are roughly $8 \times 10^{11}$ graphs,
so the fraction of non-inequivalent produced random graphs is expected
to be less than $10^{-4}$.

Analogously, we have examined $100$ millions random $4$-regular
biconnected graphs with $21$ vertices.  We have checked that
Eq. (\ref{hlk}) is satisfied for $k=3$.  From Table \ref{tabdeg4ne},
we estimate that for $v=21$ there are roughly $2\times 10^{12}$
graphs, so the fraction of equivalent random graphs produced by the
generator, is expected to be less than $10^{-4}$.

In all the tests performed on regular graphs (over more than $300$
million graphs) no violations of Eq. (\ref{hlk}) for $k=3$ are observed.

As a further comment, we observe that there are some similarities
between the graph positivity property[\onlinecite{bfppos}] and
Eq. (\ref{Delta0}): in both cases the violating graphs have an average
order of the automorphism groups which is several times larger than the
average over all the graphs with the same number of vertices $v$.
This property is observed also in the case of Eq. (\ref{hlk}), but to a
lesser extent.

\section{Conclusions}
We have observed that the coefficients of the virial expansions of the
MD model computed up to now satisfy Eq. (\ref{m2}) on all infinite
regular lattices. In particular the first $20$ coefficients of the
virial expansion satisfy Eq. (\ref{m2}) for the hypercubic infinite
lattices of any dimension $d$.  This led us to {\it conjecture} that
the virial coefficients are {\it all} positive for any infinite
regular lattice model, and to the stricter {\it conjecture} that they
all satisfy Eq. (\ref{m2}).

Using a simple relation between the virial coefficients and the
coefficients of the series for the dimer entropy, the {\it conjecture}
on the positivity of the virial coefficients led us to test the
validity of the bounds in Eq. (\ref{Delta0}) for the finite lattice
graphs and also for the finite graphs which somehow generalize them,
namely the biconnected regular graphs.

For $k=2$ these bounds  follow from the inequality Eq. (\ref{hl})
proved by Heilmann and Lieb.

We have shown that this inequality leads to {\it rigorous} upper
bounds for the number of matchings $N(i)$ improving those known up to
now for regular graphs[\onlinecite{friedbnd}] in the region of low
dimer density, and for general graphs. In the latter case, we derive an
upper bound for the matching matching polynomial improving the one in
[\onlinecite{carroll}].

The tests on lattice graphs support the validity of the virial
positivity {\it conjecture}.

Our tests on a large class of regular graphs and lattice graphs
also support the {\it conjecture} that the frequency of violations of 
Eq. (\ref{Delta0}) tends to zero as $v \to \infty$. 

We have proposed the more general bounds Eq. (\ref{hlk}) corresponding
to the stricter {\it conjecture} that Eq. (\ref{m2}) is valid for
infinite regular lattices .

Our tests of the bounds Eq. (\ref{hlk}) on the regular graphs 
give results similar
 to those obtained for the bounds Eq. (\ref{Delta0}), although with
more violations.

The tests of Eq. (\ref{hlk}) on lattice graphs show many more
violations than for the bounds Eq. (\ref{Delta0}); there are however
bands with $L_x \approx L_y$ in the 2d case and with $L_x \approx L_y
\approx L_z$ for the cubic slabs, for which there are no violations.
Extrapolating the behavior in these regions to the limit of large
lattices we get some indication that there are no violations of bounds
Eq. (\ref{hlk}) and the inequality Eq. (\ref{m2}) in the infinite
lattice limit, but this indication is not as strong as in the case of
the bounds Eq. (\ref{Delta0}) and virial positivity.

In all tests (carried out over more than $300$ million regular
 graphs), we found no violations of the bounds Eq. (\ref{hlk}) for
 $k=3$.  It would be interesting to know whether the bounds
 Eq. (\ref{hlk}) for $k=3$ are satisfied by all regular biconnected
 graphs.

 We have discussed upper bounds that could be proved if Eq. (\ref{hlk})
 held for $k=3$: they are tighter than those we proved based on the
 truth of the $k=2$ case.

\section{Appendix A: Virial coefficients in $d=2,3$}

The virial coefficients $m_k$ in $d=2,3,4$ can be obtained from the
entropy coefficients $a_k$ computed in [\onlinecite{bfp}] through order $24$;
they satisfy the bound Eq.(\ref{m2}).
In Table \ref{virial123} we give the virial coefficients in $d=2,3$.

\begin{table}[ht]
\caption{ Virial coefficients $m_k$ in $d=2,3$ up to order $24$.}
\begin{tabular}{|c|c|c|}
 \hline
$k$   & $d=2$  & $d=3$\\
 \hline
2 & 7/16 & 11/24\\
3 & 31/96 & 71/216\\
4 & 121/512 & 419/1728\\
5 & 471/2560 & 31/160\\
6 & 1867/12288 & 15031/93312\\
7 & 7435/57344 & 89951/653184\\
8 & 29477/262144 & 539963/4478976\\
9 & 116383/1179648 & 3244127/30233088\\
10 & 459517/5242880 & 19482611/201553920\\
11 & 1821051/23068672 & 116960471/1330255872\\
12 & 7255915/100663296 & 702028151/8707129344\\
13 & 29063919/436207616 & 1404544085/18865446912\\
14 & 16697149/268435456 & 8431212005/121899810816\\
15 & 157001097/2684354560 & 151861682911/2350924922880\\
16 & 1898046421/34359738368 & 911893249819/15045919506432\\
17 & 7634823999/146028888064 & 5475822286271/95917736853504\\
18 & 30619146937/618475290624 & 32879301057475/609359740010496\\
19 & 122399296903/2611340115968 & 197399821995527/3859278353399808\\
20 & 488028559661/10995116277760 & 395010395194633/8124796533473280\\
21 & 1943383170991/46179488366592 & 2371215487550117/51186218160881664\\
22 & 7740357251909/193514046488576 & 3882036150042265/87747802561511424\\
23 & 30871937807467/809240558043136 & 256340050245714583/6054598376744288256\\
24 & 123369796036139/3377699720527872 & 1538798075750480935/37907050706572935168\\

 \hline
\end{tabular}
\label{virial23}
\end{table}

\clearpage

\section{Appendix B: Proof of Eqs.(\ref{Delta0}, \ref{hlk}) for some 
classes of graphs}

Let
\begin{equation}
    A(a, v, k) = \Delta^k {\rm \ln}(N(i)i!((n-i)!)^a)
\nonumber
\end{equation}
with $n = [\frac{v}{2}]$ and $k \ge 2$.
The bound Eq. (\ref{Delta0}) is $A(0, v, k) \le 0$, the bound
Eq. (\ref{hlk}) is $A(1, v, k) \le 0$.

In the case of polygons we have
\begin{equation}
N(i) = \frac{v}{v-i}\binom{v-i}{i}
\nonumber
\end{equation}
 and thus 
\begin{equation}
A(a, v, k) =  \Delta^{k-1} \left( {\rm \ln}(1 - \frac{2i}{v}) +
{\rm \ln}(1 - \frac{2i+1}{v}) - {\rm \ln}(1 - \frac{i + 1}{v})
- a {\rm \ln}(1 - \frac{i}{n}) \right)
\nonumber
\end{equation}
For $a=0$ and all $v$ or $a=1$ and $v$ even,
\begin{equation}
A(a, v, k) = \Delta^{k-1}\sum_{h=1}\frac{1}{jv^j}\left(
(i+1)^j - (2i+1)^j - (1-a)2^j \right) < 0
\nonumber
\end{equation}

In the case of the complete graphs $K_{v}$, we have  
\begin{equation}
N(i) = \frac{v!}{(v-2i)!i!2^i}
\nonumber
\end{equation}
and thus 
\begin{equation}
A = \Delta^{k-1}\left(
{\rm \ln}(1 - \frac{2i}{v}) + {\rm \ln}(1 - \frac{2i+1}{v}) 
- a {\rm \ln}(1 - \frac{i}{n})\right) 
\nonumber
\end{equation}
For $a=0$ and all $v$ or $a=1$ and even $v$  one gets
\begin{equation}
A = \Delta^{k-1}\sum_{j=1}\frac{1}{jv^j}\left( -(1-a)(2i)^j - (2i+1)^j \right)
< 0
\nonumber
\end{equation}

In the case of the complete bipartite graphs $K_{n,n}$, we have
\begin{equation}
N(i) = \binom{n}{i}^2i!
\nonumber
\end{equation}
and thus 
\begin{equation}
A(a, v, k) =  (2-a) \Delta^{k-1} {\rm \ln}(1 - \frac{i}{n}) < 0
\nonumber
\end{equation}

Friedland, Krop and  Markstr$\ddot{ \rm o}$m[\onlinecite{FKM}] have computed 
an average distribution for
$ N(i)$ in the case of random regular bipartite graphs
\begin{equation}
N(i) = \frac{\binom{n}{i}^2 r^{2i} i!(r n - i)!}{(r n)!}
\label{Nfkm}
\end{equation}
and thus
\begin{equation}
A = \Delta^{k-1} (
(2-a){\rm \ln}(1 - \frac{i}{n}) - {\rm \ln}(1 - \frac{i}{rn})) =
\Delta^{k-1}\sum_{J=1}\frac{1}{j}\left(-(2-a)(\frac{i}{n})^j +
(\frac{i}{rn})^j \right) < 0
\nonumber
\end{equation}

The easily derived bipartite mean-field approximation 
\begin{equation}
N(i) = \binom{n}{i}^2 i! \left(\frac{r}{n}\right)^i
\nonumber
\end{equation}
satisfies the bounds; the proof is similar to that for $K_{n,n}$.

\section{Appendix C}
Consider the bounds

\begin{equation}
\Delta^k g(i) \le 0
\label{g0}
\end{equation}
valid  for some $k \ge 1$ and for $i \ge i_0$.
Summing from $i_0$ to $i-1$ one gets
\begin{equation}
\Delta^{k-1} g(i) \le \Delta^{k-1} g(i_0)
\label{g1}
\end{equation}
Summing again and using
\begin{equation}
\sum_{k=1}^{n} \binom{k}{m} = \binom{n+1}{m+1}
\nonumber
\end{equation}
we can prove by induction that, for $h \le k$
\begin{equation}
\Delta^{k-h}g(i) \le \sum_{t=0}^{h-1} \binom{i-i_0}{t} \Delta^{k-h+t}g(i_0)
\label{g3}
\end{equation}

In the case $h=k$ Eq. (\ref{g3}) gives
\begin{equation}
g(i) \le  \sum_{t=0}^{k-1}\binom{i-i_0}{t}\Delta^tg(i_0)
\label{g4}
\end{equation}
for $i \ge i_0$.

If a graph satisfies the bound Eq. (\ref{hlk}) for a given $k$, then
Eq. (\ref{g4}) with $g(i) = {\rm \ln} P(i)$, where $P$ is defined in
Eq. (\ref{Pi}), gives an upper bound for $N(i)$.

\end{document}